\documentclass[runningheads]{llncs}
\usepackage[T1]{fontenc}
\usepackage{graphicx}
\usepackage{soul}
\usepackage{url}
\usepackage{mathtools}
\usepackage{subcaption}
\usepackage{booktabs}
\usepackage{hyperref}
\usepackage{listings}
\usepackage{xcolor}
\usepackage{color}
\lstdefinestyle{promptstyle}{
  basicstyle=\ttfamily\footnotesize,
  breaklines=true,
  breakatwhitespace=false,
  columns=fullflexible,
  keepspaces=true,
  showstringspaces=false,
  frame=single,
  framesep=4pt,
  rulecolor=\color{gray!50},
  backgroundcolor=\color{gray!5},
  escapeinside={(*@}{@*)},
}

\begin{document}
%

\title{Evaluating RAG for French immigration law: a benchmark and baseline study}
\titlerunning{Evaluating RAG for French immigration law}
%
\author{Annia Abtout\thanks{These authors contributed equally.}\orcidID{0009-0006-2377-1926} \and Julien Delaunay$^*$\orcidID{0009-0001-9247-5745} \and Monika Ewa Rakoczy\orcidID{0000-0001-5249-5509}}
\authorrunning{A. Abtout et al.}
%
\institute{Talan, Paris, France \\
\email{first\_name.name@talan.com}}
\maketitle              

\begin{abstract}
International recruitment in France requires navigating a layered legal framework absent from existing legal AI benchmarks. We present a publicly available benchmark and first comparative evaluation for this domain, covering permit-type recommendation, required-document retrieval, and legal citation coverage. Comparing a parametric LLM baseline against dense retrieval augmentation at two model scales (Qwen3.5-9B and -27B) on 52 annotated synthetic profiles, we find that retrieval improves administrative guidance at both scales, most notably permit-type accuracy. Our results confirm that retrieval grounding is important for more reliable administrative guidance in this domain, and motivate further investigation of hybrid retrieval strategies.

\end{abstract}

\section{Introduction}

International recruitment requires navigating administrative procedures that determine whether and under which conditions a foreign national may legally work in a given country. In France, Human Resource (HR) practitioners must identify the appropriate residence permit, determine the supporting documents required by the administration, and ensure that these recommendations are consistent with the applicable regulatory framework. This process draws on multiple sources, including immigration law, labour regulations, bilateral agreements, and administrative guidance.

Recent advances in Large Language Models (LLMs) have stimulated growing interest in legal and regulatory AI applications, supported by benchmarks such as LegalBench~\cite{guha2023legalbench}, LawBench~\cite{fei2024lawbench}, and LexEval~\cite{li2024lexeval}. However, to our knowledge, there is no publicly available benchmarks targeting administrative guidance for international recruitment in the French immigration context. Moreover, the task differs from many existing legal benchmarks as a system must produce a structured recommendation combining an administrative decision, the associated procedural requirements, and sufficient legal grounding to support the recommendation.

In this work, we introduce a publicly available benchmark for French immigration administrative procedures in the context of international recruitment. The benchmark contains 52 annotated synthetic applicant profiles and supports evaluation along three dimensions: residence permit type, required documents and legal basis. Using this benchmark, we conduct a first comparison, at two model scales, between a parametric LLM baseline and a Retrieval-Augmented Generation (RAG) \cite{lewis2020retrieval} approach. 
Code, benchmark, and annotation guidelines are available through this repository.\footnote{\url{https://github.com/jdelaunay/foreign_workers_recruitment_dataset}}

\section{Related work}
\paragraph{Legal AI and LLMs Limitations.}
LLMs have demonstrated impressive capabilities in various areas but systematic studies reveal that GPT-4 \cite{achiam2023gpt} hallucinates in at least 58\% of cases when queried about U.S. case law \cite{dahl2024large}, generating what researchers term ``large legal fictions''. These models exhibit sycophancy bias, accepting erroneous legal premises rather than correcting them \cite{keisha2025all,savelka2023explaining}. This unreliability has motivated the development of more constrained architectures.
\paragraph{Legal Benchmarks and RAG Validation.}
Recent benchmarks \cite{guha2023legalbench,li2024lexeval,fei2024lawbench,pipitone2024legalbench} demonstrate that RAG outperforms both generic and fine-tuned LLMs in legal reasoning, with LexEval showing that specialized models excel at memorization but fail at logical inference. However, retrieval quality, particularly chunking and context management, remains the primary performance bottleneck. Existing benchmarks predominantly cover common law systems, leaving civil law jurisdictions underexplored. Adaptive systems like HyPA-RAG~\cite{kalra2024hypa} and LexRAG~\cite{li2025lexrag} address legal complexity by dynamically adjusting retrieval strategies, though at substantial computational cost. To the best of our knowledge, the intersection of legal reasoning and administrative guidance in the French immigration context remains unexplored, despite its practical importance for international recruitment.

\section{Benchmark construction}
\subsection{Legal framework} \label{sec:legal_framework}
The recruitment of foreign workers in France operates under a hierarchical legal framework combining immigration law, labour regulations, bilateral agreements, and administrative procedures. The Code on the Entry and Residence of Foreign Nationals and the Right of Asylum (CESEDA) defines the main residence permit categories and eligibility conditions. For nationals of countries covered by bilateral agreements, these agreements take precedence over CESEDA provisions. The most prominent example is the Franco–Algerian Agreement, which establishes a distinct legal regime for Algerian nationals. Labour market access conditions are further governed by the Employment Act (Book V, articles L5221–L5224 and associated regulatory provisions), while administrative procedures are documented through the Service-Public\footnote{\url{https://www.service-public.gouv.fr}} portal and, in some cases, supplemented by prefecture-specific requirements.
As a consequence, administrative guidance requires reasoning across multiple legal sources whose applicability depends on applicant characteristics. This distinguishes the task from standard legal question answering and motivates the construction of a dedicated benchmark.

\subsection{Task description}
While traditional legal benchmarks focus on question answering or legal reasoning in isolation, immigration guidance requires the simultaneous production of several complementary outputs. Given a structured applicant profile, the system must identify the appropriate residence permit category, determine the supporting administrative documentation, and provide the legal references justifying the recommendation.
Accordingly, each benchmark instance contains three target components: a permit recommendation composed of four attributes (permit category, statutory mention, renewal status, and visa requirement), the set of documents required to support the application and a set of legal references corresponding to the applicable legal basis. This multi-output formulation reflects the operational needs of HR practitioners, who must identify the correct permit and understand the associated administrative procedure and its legal justification.

\subsection{Dataset construction}

The benchmark combines synthetic applicant profiles and annotations of the corresponding residence permit application process.
Applicant profiles were generated using a rule-based simulation process calibrated on publicly available demographic statistics from the French National Institute of Statistics and Economic Studies (INSEE). We adopted synthetic profiles because real immigration cases are difficult to share due to privacy and confidentiality constraints. Also, real-world datasets often under-represent rare permit categories and legal situations. Finally, synthetic generation allows controlled coverage of the main legal regimes encountered in international recruitment. The resulting profiles cover different nationalities, residence situations, employment conditions, and permit categories. Detailed generation parameters are provided in Supp. \ref{sec:profile_generation}.
To facilitate annotation, we employed a RAG-augmented Qwen3.5-27B \cite{qwen3.5} over a curated legal corpus (Supp.~\ref{sec:legal_corpus}) to generate candidate recommendations, which annotators then validated or corrected using the INCEpTION annotation tool \cite{klie-etal-2018-inception}.
The annotation campaign involved six annotators with varying levels of familiarity with French immigration procedures and an HR specialist in international recruitment acting as domain expert. Annotators independently produced permit recommendations, required documents lists, and legal references following a shared annotation guideline. To ensure consistency across annotations, annotators were instructed to rely on a predefined set of authoritative legal sources covering immigration law, labour regulations, bilateral agreements, and administrative guidance. Five calibration profiles were first annotated by all participants and the HR expert to refine the protocol. The remaining profiles were double-annotated and disagreements were adjudicated by a third annotator. This process yielded 52 curated benchmark instances.
To assess benchmark reliability, we measured inter-annotator agreement on all target dimensions. Agreement was moderate for permit and documents and higher for legal basis annotations, with most disagreements arising from optional residence permits for EU citizens and the distinction between permit renewal and permit upgrade. Detailed agreement analyses are provided in Supp. \ref{sec:iaa}.
\section{Experiments}
\subsection{Evaluation settings}
We evaluate the effect of retrieval on recommendation quality across two model scales, Qwen3.5-9B and Qwen3.5-27B \cite{qwen3.5}. For each model we compare \textit{LLM-only} which receives the applicant profile and a structured prompt, without access to any external knowledge and serves as a parametric baseline with \textit{Dense RAG} which augments the same model with retrieved legal context drawn from a vectorized legal corpus through MediaTech \footnote{MediaTech is an open-source platform providing pre-processed and embedded public datasets for AI applications : https://beta.gouv.fr/startups/media-tech.html }. The corpus combines resources from all the relevant legal sources (see Supp. \ref{sec:legal_corpus}). Documents are segmented into semantically coherent chunks corresponding to articles, sections, or procedural content blocks and are represented using pre-computed BGE-M3 embeddings \cite{chen-etal-2024-m3}. Given an applicant profile, dense vector search retrieves the top-$k$ most relevant chunks, which are then injected into the LLM prompt.
All four conditions share the same prompt template (see Supp. \ref{app:prompt}) and legal corpus, so that observed differences are attributable to the two controlled factors : the presence or absence of retrieval and model scale. Experiments are conducted on the annotated subset of our benchmark, which covers the main nationality groups and permit categories introduced in Supp. \ref{sec:profile_generation} and ~\ref{sec:dataset_stats}. 

\subsection{Evaluation metrics}
\label{sec:metrics}
The benchmark evaluates three aspects of french administrative guidance so we define one task-specific metric for each output dimension.
\subsubsection*{Permit-type.}
Each recommendation is decomposed into four components:
\texttt{is\_renewal}, \texttt{permit\_type} (16-category closed taxonomy, Table~\ref{tab:permit_taxonomy} in Supp. \ref{app:taxonomy}), \texttt{mention} (e.g.\ \emph{skilled worker}) and \texttt{visa\_required}. Each component is evaluated as a binary match against the gold standard. Full permit match requires all four components to be simultaneously correct. We additionally report \texttt{mention*} as mention accuracy conditioned on correct \texttt{permit\_type}, to isolate intra-category errors. Normalisation and matching rules are described in Supp.~\ref{app:normalisation}.

\subsubsection*{Documents.}

Document requirements are evaluated only on full permit match profiles, which isolate downstream document prediction from upstream permit-classification errors. Matching combines rule-based normalisation of frequent administrative variants with embedding-based semantic alignment. We report precision, recall and F1 score. See Supp.~\ref{app:doc-matching} for matching procedure.

\subsubsection*{Legal citation.}

Legal grounding is evaluated through a recall-oriented coverage
metric. Since the gold standard contains a minimal set of supporting
references, precision would be
difficult to interpret. Predicted citations are first mapped to
canonical representations (CESEDA, Service Public or bilateral agreements), after which coverage is
computed as the proportion of gold references recovered by the
system. This metric is also restricted to full permit match profiles.

\section{Results}\label{results}
\subsubsection*{Permit type recommendation accuracy}

\begin{table}[t]
\centering
\caption{Component-level permit accuracy for Qwen-27B and Qwen-9B.
$\Delta$ is the absolute difference between Dense RAG and LLM-alone.
\emph{mention$^*$} is conditional on \emph{permit\_type} being correct.}
\label{tab:permit_accuracy}
\begin{tabular}{lcccccc}
\toprule
& \multicolumn{3}{c}{\textbf{Qwen-27B}} & \multicolumn{3}{c}{\textbf{Qwen-9B}} \\
\cmidrule(lr){2-4} \cmidrule(lr){5-7}
\textbf{Component} & \textbf{LLM-alone} & \textbf{Dense RAG} & \boldmath{$\Delta$}
& \textbf{LLM-alone} & \textbf{Dense RAG} & \boldmath{$\Delta$} \\
\midrule
\texttt{is\_renewal}         & 88.5\% & 82.7\% & $-$5.8 pts
                             & 82.7\% & 82.7\% & $\pm$0.0 pts \\
\texttt{permit\_type}        & 50.0\% & 75.0\% & $+$25.0 pts
                             & 30.8\% & 57.7\% & $+$26.9 pts \\
\texttt{mention}$^*$         & 84.6\% & 94.9\% & $+$10.3 pts
                             & 81.2\% & 90.0\% & $+$8.8 pts \\
\texttt{visa\_required}      & 53.8\% & 55.8\% & $+$1.9 pts
                             & 53.8\% & 53.8\% & $\pm$0.0 pts \\
\midrule
\texttt{full\_permit\_match} & 25.0\% & 34.6\% & $+$9.6 pts
                             & 9.6\% & 25.0\% & $+$15.4 pts \\
\bottomrule
\end{tabular}
\end{table}


Table~\ref{tab:permit_accuracy} reports component-level permit accuracy across the 52 benchmark profiles. Dense RAG improves permit recommendation at both scales:
\texttt{permit\_type} accuracy rises from 50.0\% to 75.0\% for Qwen-27B and from
30.8\% to 57.7\% for Qwen-9B, and \texttt{full\_permit\_match} from 25.0\% to
34.6\% and from 9.6\% to 25.0\% respectively. The relative gain is at least as
large for the smaller model, whose weaker parametric baseline points to a stronger reliance on retrieval.
Retrieval seems to narrow but not close the gap with the larger model. Conditional
mention accuracy (\texttt{mention}$^*$) is high and improves at both scales which indicates that,
once the permit family is correct, the appropriate mention is generally
recovered.

Error analysis (Table~\ref{tab:error_analysis}, Supp.~\ref{app:permit_match})
indicates that retrieval primarily improves identification of the applicable
legal regime: the LLM-alone baselines frequently confuse permit families
governed by different frameworks and Dense
RAG substantially reduces these errors. The same error families appear at both
scales, with the 9B model defaulting more often to the most common categories for workers
(\emph{employee}, Talent).

By contrast, retrieval has little effect on \texttt{visa\_required}, which is
essentially unchanged across both conditions and both scales. Indeed, most residual
errors correspond to omitted visas. Since visa requirements were not explicitly
emphasised in the prompt, these should not be read solely as retrieval failures. 
\subsubsection*{Document requirement matching}
\label{sec:results-docs}
Table~\ref{tab:doc_f1} reports document-set matching under the strict full permit match filter. Dense RAG substantially improves document
prediction at both scales, raising F1 from 23.9\% to 45.3\% for Qwen-27B and from 23.5\% to 36.4\% for
Qwen-9B with consistent gains in
precision and recall. The gain appears smaller for the 9B model, which appears to
exploit retrieved context less effectively. Because evaluation is restricted to
profiles with correct permit recommendations (a small subset), these scores should be read qualitatively.

Error analysis (Supp.~\ref{sec:doc-errors}) suggests that retrieval primarily
reduces hallucinated administrative requirements while improving recovery of
permit-specific supporting documents. However, several procedural requirements
remain frequently omitted in both conditions, indicating that retrieval alone is
insufficient to guarantee exhaustive procedural guidance.

\subsubsection*{Legal citation quality}
\label{sec:results-citations}
Dense RAG improves legal grounding for Qwen-27B by raising citation
coverage from 7.7\% to 46.6\%  (Table~\ref{tab:citation_coverage},
Supp.~\ref{sec:legal-errors}), with references recovered from all source types. For
Qwen-9B the corresponding improvement is much smaller (10.0\% to 13.1\%), with
almost no Service Public or bilateral references recovered. We stress, however,
that the 9B citation figures rest on a very small conditional subset ($n=5$ and
$n=13$ profiles) so this contrast
should be read as a tentative indication that citation grounding may be more
scale-dependent than permit classification, not as an established effect.
Coverage remains incomplete even for Qwen-27B, and all scores should be treated
as lower bounds, since the gold standard contains only a minimal reference set.
\section{Discussion and conclusion}
The benchmark campaign exposed recurring ambiguities, including visa requirements administrative vocabulary, and legal citation exhaustivity. One field definition was revised mid-campaign (degree \textit{obtained} became \textit{recognised} in France) requiring retrospective adjudication. Such issues are inherent to first-round benchmark construction in a domain where legal practice itself admits variability. Even with structured applicant profiles and a simplified representation focused on permit eligibility, documents and legal references, recommendations remain sensitive to profile attributes. The challenges observed during annotation and evaluation therefore reflect the complexity of the underlying administrative framework.

The results suggest that this complexity remains challenging for current AI systems. Retrieval grounding improves permit classification at both model scales indicating that retrieval partly compensates for limited parametric knowledge. Nonetheless, errors persist despite a constrained corpus and the benefit of retrieval is less uniform on the downstream dimensions: gains on
document and citation recovery are smaller for the 9B model, suggesting that exploiting retrieved context for detailed legal grounding is more scale-dependent than permit classification. As our evaluation still covers a single model family, broader cross-family validation remains future work. Our findings are otherwise consistent with prior work on legal and administrative reasoning.

Several limitations should also be acknowledged. The benchmark contains only 52 annotated profiles and the documents and citation evaluations are conditioned on correct permit classification, yielding small subsets so cross-scale differences on these two dimensions should be read as indicative only. Legal citation coverage should be interpreted as a conservative lower bound, since annotators provided only a minimal set of references. Also, using the same Qwen model family for generation and evaluation might have introduced a self-preference bias \cite{panickssery2024llm}. Future work should broaden the range of model families and scales and investigate richer evaluation protocols for legal grounding. Hence, this work informs two next steps. On the benchmark side, a second annotation campaign targeting at least 200 profiles is planned, with a revised protocol addressing identified ambiguities. On the system side, persistent under-retrieval points to hybrid retrieval
(HyPA-RAG~\cite{kalra2024hypa}), following the state of the art.

\begin{credits}
\subsubsection{\ackname} We express our gratitude to Khaouthar MOUSSAOUI, Elodie WIKSZAK, Thierry SCHULLER, Pauline LALLEMENT and Tsamba MBEMBO for their invaluable assistance in the annotation process. We would also like to thank Laurent CERVONI, head of the team, for his support and guidance throughout this project. We are also grateful to Florence REAL for identifying the initial use case within the HR department that gave rise to this project.
\subsubsection{\discintname}
The authors have no competing interests to declare that are
relevant to the content of this article.
\end{credits}
%
%

\bibliographystyle{splncs04}
\bibliography{samplepaper}

\appendix
\section*{Supplementary material}
\section{Benchmark building}
\subsection{Profiles generation} \label{sec:profile_generation}
The objective of profile generation was not to reproduce the exact distribution of immigration cases observed in practice, but rather to obtain a benchmark covering the principal legal regimes encountered during international recruitment while preserving demographic plausibility.

To create a realistic evaluation dataset, we developed a demographic simulation system that generates synthetic applicant profiles reflecting France's immigrant population distribution based on the French National Institute of Statistics and Economic Studies (INSEE)\footnote{The INSEE collects, produces, analyses and disseminates information on the French economy and society} empirical data.
Nationality distribution follows observed immigration patterns in 2024\footnote{\url{https://www.insee.fr/fr/statistiques/3633212}} through weighted sampling across four regions: Africa (42.2\%, led by Algeria 12.4\%, Morocco 11.7\%), Europe (26.5\%, led by Portugal 7.3\%), Asia (14.3\%, led by Turkey 3.4\%), and the Americas (4.0\%). Two-thirds of profiles indicate current French residence. Geographic locations are sampled from GeoNames\footnote{\url{https://www.geonames.org}} with population-based weights.
Educational attainment skews toward higher qualifications (Master's 25\%, Bachelor's 20\%), notably to enable talent card eligibility for some profiles. Professional status heavily favors permanent employment (75\%). Proposed salaries follow a right-skewed distribution (Skew-normal, $\alpha$=3) centered at €35,000 with a €21,876.40 minimum threshold, to allow high salaries which can allow specific residence permits like the talent card.
Current residence permits are assigned based on time in France: recent arrivals (0–1 years) predominantly hold temporary residence permit (55\%), mid-term residents (1–6 years) hold multi-year residence permit, and long-term residents (7+ years) primarily hold 10-year residence permits (70\%). Algerian nationals receive certificates of residence under the Franco-Algerian Agreement. Ninety percent of permits approach expiration (1–4 months remaining), simulating typical renewal scenarios.

\subsection{Legal Corpus Construction}
\label{sec:legal_corpus}
Because retrieval quality is inherently bounded by corpus quality, particular attention was devoted to source selection and legal validity filtering.

The legal corpus underlying our framework aggregates four heterogeneous normative sources, each subject to source-specific preprocessing and filtering prior to indexing. This filtering stage is a deliberate architectural choice: indexing abrogated or thematically irrelevant provisions would expose the retrieval pipeline to legally incorrect recommendations regardless of retrieval quality, directly undermining the traceability guarantees.

The CESEDA corpus, built from the French consolidated legislation dataset\footnote{\url{https://huggingface.co/datasets/AgentPublic/legi}}, retains only articles whose status is marked
\textit{in force}, excluding abrogated provisions that remain present in the raw dataset. The same validity filter is applied to the Employment Act, constructed from the same corpus, which is further restricted to articles targeting Book~V, Title~II, Chapter~II, Sections~1--4, the statutory provisions governing labor market access conditions for foreign nationals (articles L5221--L5224 and their regulatory counterparts). This targeted restriction reduces index noise from unrelated labor law provisions while preserving full coverage of the relevant perimeter.

The bilateral agreements corpus is constructed by merging two source datasets (\texttt{legi\_decret} and \texttt{legi\_decret\_loi}) from the French consolidated legislation dataset, filtered to in-force texts whose titles simultaneously reference the French Republic, a legal instrument type (\textit{agreement}, \textit{convention}, \textit{joint statement}), and a thematic scope related to persons, residence, or employment. This compound filter operationalizes the legal hierarchy described in Section~\ref{sec:legal_framework}, ensuring that only instruments with direct bearing on immigration and labor rights are indexed.

Finally, the Service-Public corpus\footnote{\url{https://huggingface.co/datasets/AgentPublic/service-public}} is indexed without thematic filtering, as its procedural scope (application workflows, required documentation, and administrative deadlines) is inherently bounded to the immigration domain by source selection.

\subsection{Permit Type Taxonomy}
\label{app:taxonomy}

Table~\ref{tab:permit_taxonomy} lists the permit categories used in the closed taxonomy for \texttt{permit\_type} evaluation. Categories marked~$\dagger$ are those for which the distinction between temporary and multi-year variants carries distinct documentary requirements and is therefore evaluated independently.

\begin{table}[ht]
\centering
\caption{Permit type taxonomy. Temporary = 1~year. $\dagger$~marks categories where
the temporary/multi-year distinction carries distinct documentary requirements.}
\label{tab:permit_taxonomy}
\begin{tabular}{ll}
\toprule
\textbf{Label} & \textbf{Description} \\
\midrule
\texttt{NO\_PERMIT}          & No permit required \\
\texttt{EU\_CITIZEN}         & EU/EEA/Swiss free movement \\
\texttt{RC\_LONG\_STAY\_EU} & Long-stay EU resident card \\
\texttt{RESIDENT\_CARD}     & 10-year resident card \\
\texttt{RC\_PERMANENT}      & Permanent resident card \\
\texttt{ARC\_1Y}                 & Algerian resident certificate (1 year) \\
\texttt{ARC\_10Y}                 & Algerian resident certificate (10 years) \\
\texttt{TALENT}              & Talent Card \\
\texttt{T\_EMPLOYEE}$^\dagger$  & Temporary work permit \\
\texttt{MY\_EMPLOYEE}$^\dagger$  & Multi-year work permit \\
\texttt{T\_PFL}$^\dagger$      & Temporary private/family life \\
\texttt{MY\_PFL}$^\dagger$      & Multi-year private/family life \\
\texttt{ENTREPRENEUR}        & Self-employed / liberal profession \\
\texttt{MY\_ENTREPRENEUR}$^\dagger$ & Multi-year entrepreneur \\
\texttt{TW}                  & Temporary worker \\
\texttt{TRAINEE}           & Trainee \\
\bottomrule
\end{tabular}
\end{table}

\subsection{Inter-annotator agreement} \label{sec:iaa}
The objective of the agreement analysis is not to establish a definitive legal ground truth, but to assess whether independent annotators converge towards similar administrative recommendations under a shared protocol.

\begin{figure}[ht]
    \centering
    \begin{subfigure}[t]{0.75\linewidth}
        \includegraphics[width=\linewidth, trim=0 0 0 0.75cm, clip]{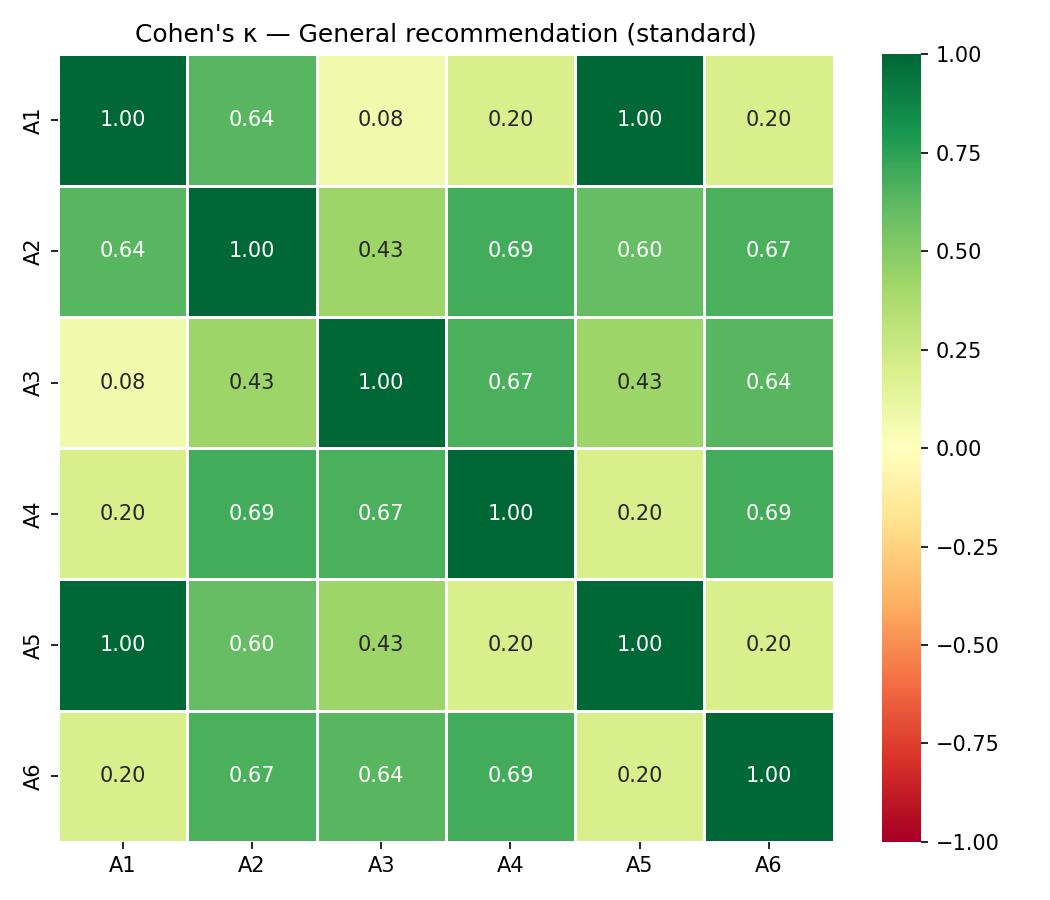}
        \caption{All errors considered}
        \label{fig:kappa_permit_rec_all}
    \end{subfigure}
    \hfill
    \begin{subfigure}[t]{0.75\linewidth}
        \includegraphics[width=\linewidth, trim=0 0 0 0.75cm, clip]{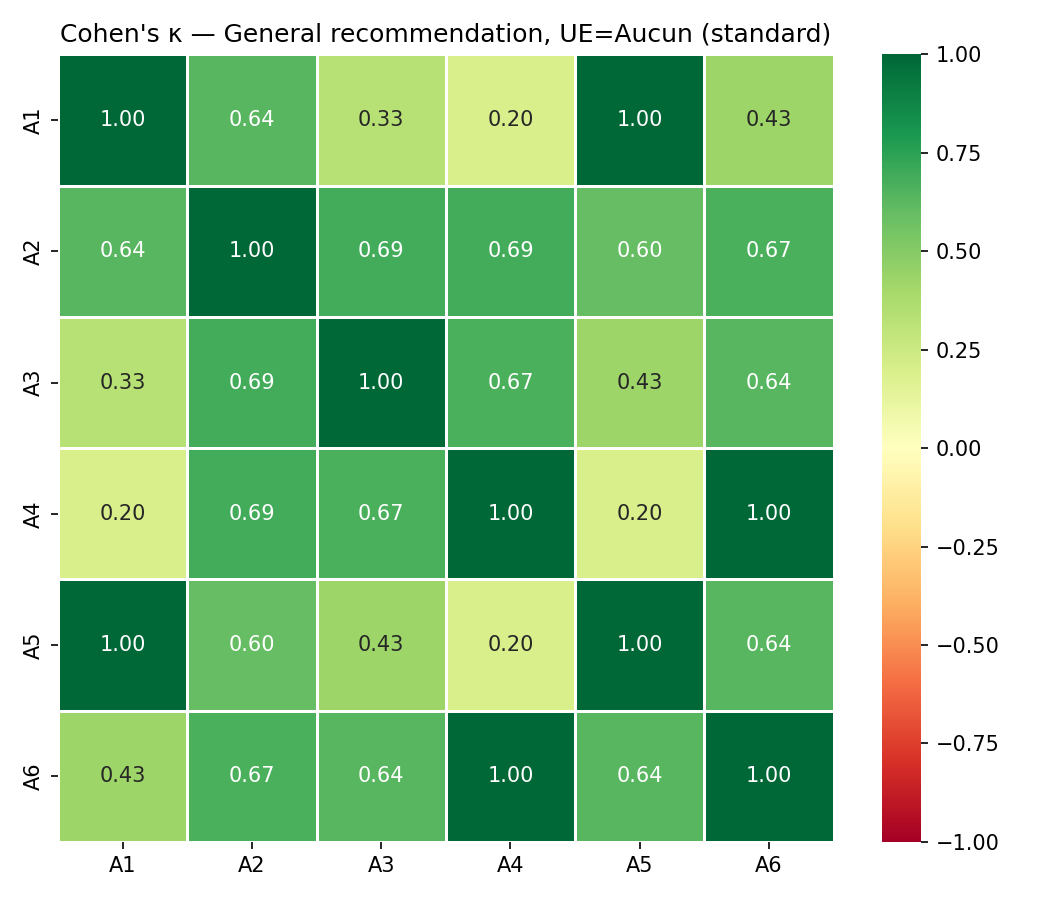}
        \caption{"No permit required" and "Residence permit for EU/EEA/Swiss citizens" considered equals}
        \label{fig:kappa_permit_rec_eu}
    \end{subfigure}
    \caption{Pairwise $\kappa$ on permit-type annotations (training profiles excluded)}
    \label{fig:kappa_permit_rec}
\end{figure}

\begin{figure}[ht]
    \centering
    \includegraphics[width=\linewidth, trim=0 0 0 0.8cm, clip]{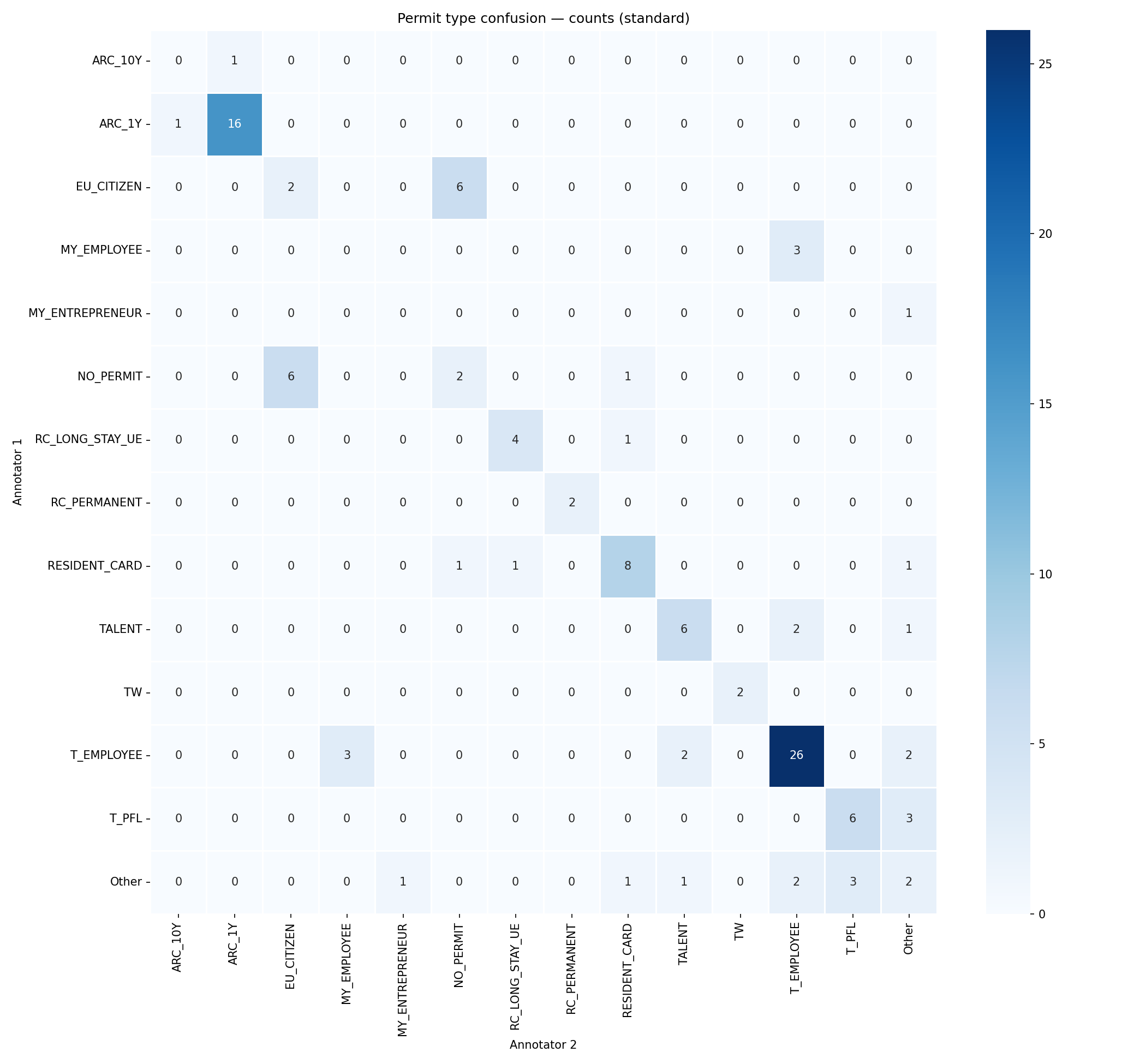}
    \caption{Confusion matrix for the permit annotation (counts). `Renewal' and visas mentions are ommited. Calibration profiles excluded.}
    \label{fig:confusion_permit_anno}
\end{figure}

Inter-annotator agreement (IAA) is reported across the three output fields using Cohen's kappa ($\kappa$) \cite{cohen1960coefficient} for permit type, and F1 score for required documents and legal basis. Figure~\ref{fig:kappa_permit_rec_all} illustrates the inherent ambiguity of permit annotation. Two systematic errors emerged: assigning a residence permit to EU/EEA/Swiss citizens despite it being optional—a guideline deviation whose impact is confirmed by the improved agreement scores in Figure~\ref{fig:kappa_permit_rec_eu} when both valid choices are treated as equivalent—and recommending permit renewals where an upgrade was available (e.g., a temporary rather than a multi-year residence permit), as can be seen on Figure \ref{fig:confusion_permit_anno}. Given the exploratory nature of this work and the limited number of annotations per pair, $\kappa$ should be interpreted as indicative only.

\begin{figure}
    \centering
    \includegraphics[width=0.75\linewidth, trim=0 0 0 0.783cm, clip]{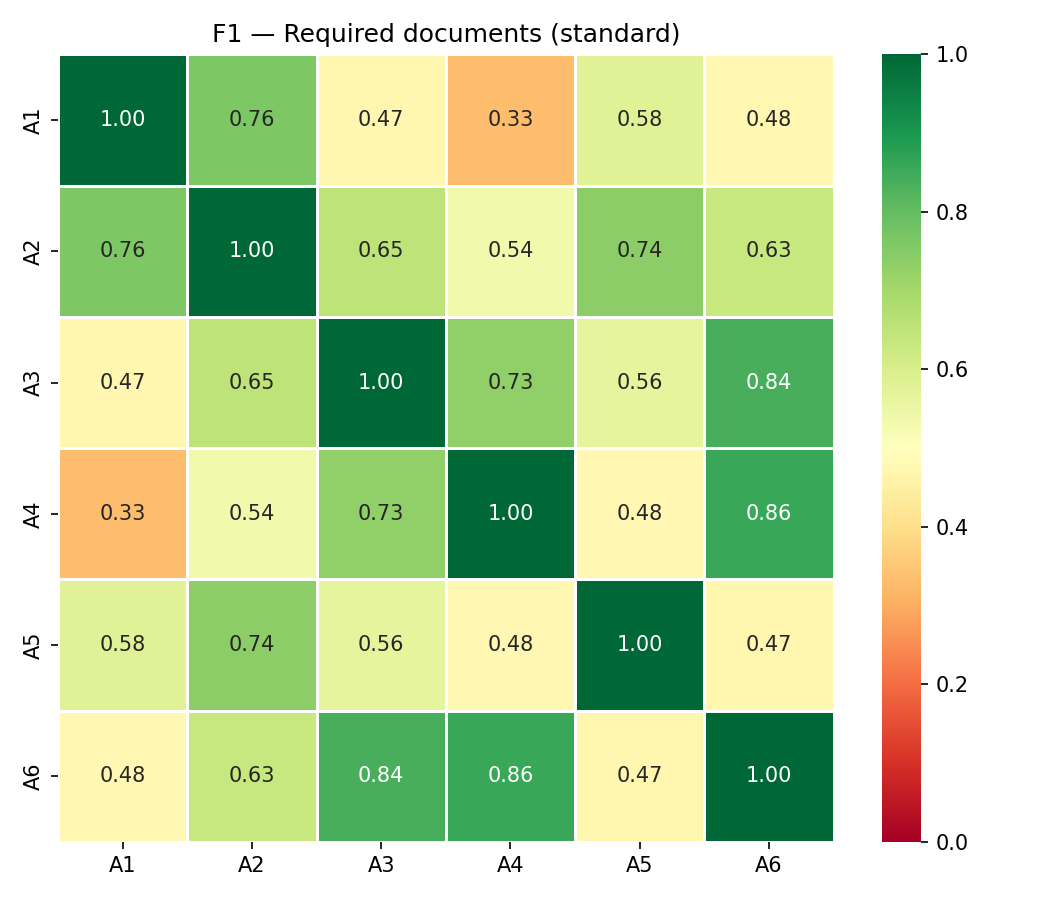}
    \caption{Pairwise F1 on required documents (calibration profiles excluded, and only profiles for which the annotators agreed on the permit-type}
    \label{fig:f1_docs}
\end{figure}

\begin{figure}
    \centering
    \includegraphics[width=0.75\linewidth, trim=0 0 0 0.775cm, clip]{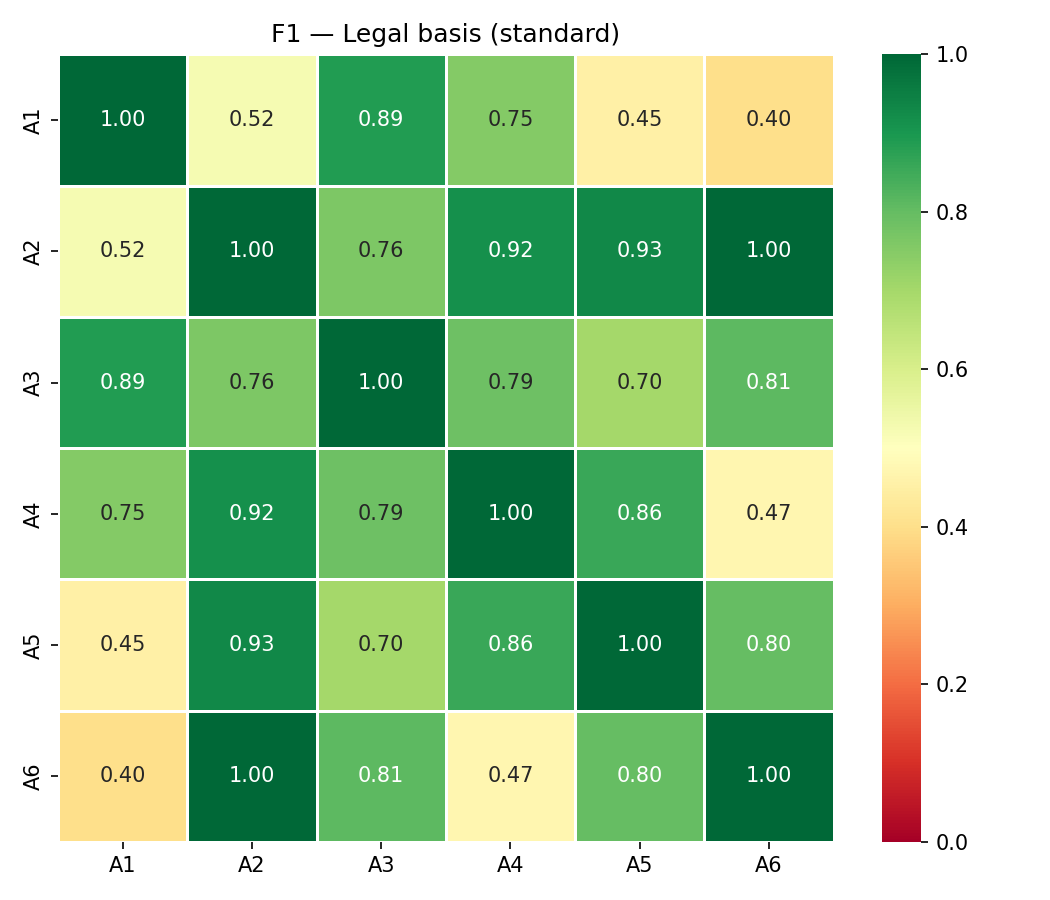}
    \caption{Pairwise F1 on legal basis (calibration profiles excluded, and only profiles for which the annotators agreed on the permit-type}
    \label{fig:f1_legal}
\end{figure}

As an exploratory contribution, the annotations covered permit type, required documents, and legal basis simultaneously. Figure~\ref{fig:f1_docs} reveals mixed results: while some documents are common across permit types (e.g., passport, passport photo), systematic errors arose around visa-required documents preceding the residence permit application, such as tax stamp of €99 or proof of qualification—sources not included in our knowledge base, requiring annotators to consult France-Visa independently. These discordances are also visible in Table \ref{tab:doc_discordances_all}. Also, most common errors concerned common documents, like the previous residence permit in case of renewal, medical certificate issued by the OFII or the signed copy of the pledge to uphold the principles of the French Republic. Using the recommandation of the LLM, while gaining time especially for the legal basis, also introduced a bias of sycophancy \cite{sharma2024towards}, which permitted some document not required to be selected by some annotators (such as employment contract or offer of employment and proof of income for instance), or incorrect fees for tax stamps.

As shown in Figure \ref{fig:f1_legal}, legal basis scores are encouraging, though ambiguities remain in the annotation guidelines themselves: it is unclear whether annotators should report the minimal legal basis strictly related to the permit type, the basis necessary to identify required documents, or an exhaustive enumeration. This warrants clarification in future work.

\begin{table}[h]
\centering
\caption{Most frequent documents involved in annotation discordances (all permit types)}
\label{tab:doc_discordances_all}
\begin{tabular}{lr}
\toprule
\textbf{Document} & \textbf{Count} \\
\midrule
Medical certificate issued by OFII & 15 \\
Valid residence permit & 13 \\
Signed commitment to respect the principles of the French Republic & 13 \\
Copy of highest diploma obtained & 12 \\
Proof of qualification & 11 \\
Long-stay visa serving as a residence permit & 11 \\
Employment contract or job offer & 10 \\
Tax stamp of €99 & 10 \\
Employment certificate & 10 \\
Proof of address (less than 6 months) & 10 \\
Proof of resources & 8 \\
Curriculum Vitae (CV) & 8 \\
Work permit issued by employer & 7 \\
Valid travel document & 6 \\
Birth certificate & 6 \\
Proof of compliance with regulatory practice conditions & 4 \\
Statutory declaration of non-polygamy & 4 \\
Tax stamp of €225 & 4 \\
Evidence supporting continuation of employment contract & 4 \\
Visa regularisation fee (€300) & 4 \\
\bottomrule
\end{tabular}
\end{table}

\subsection{Dataset statistics}
\label{sec:dataset_stats}
This section explores the statistics related to our 52 curated annotations.

\begin{figure}[ht]
    \centering
    \includegraphics[width=\linewidth,trim=0 0 0 0.9cm, clip]{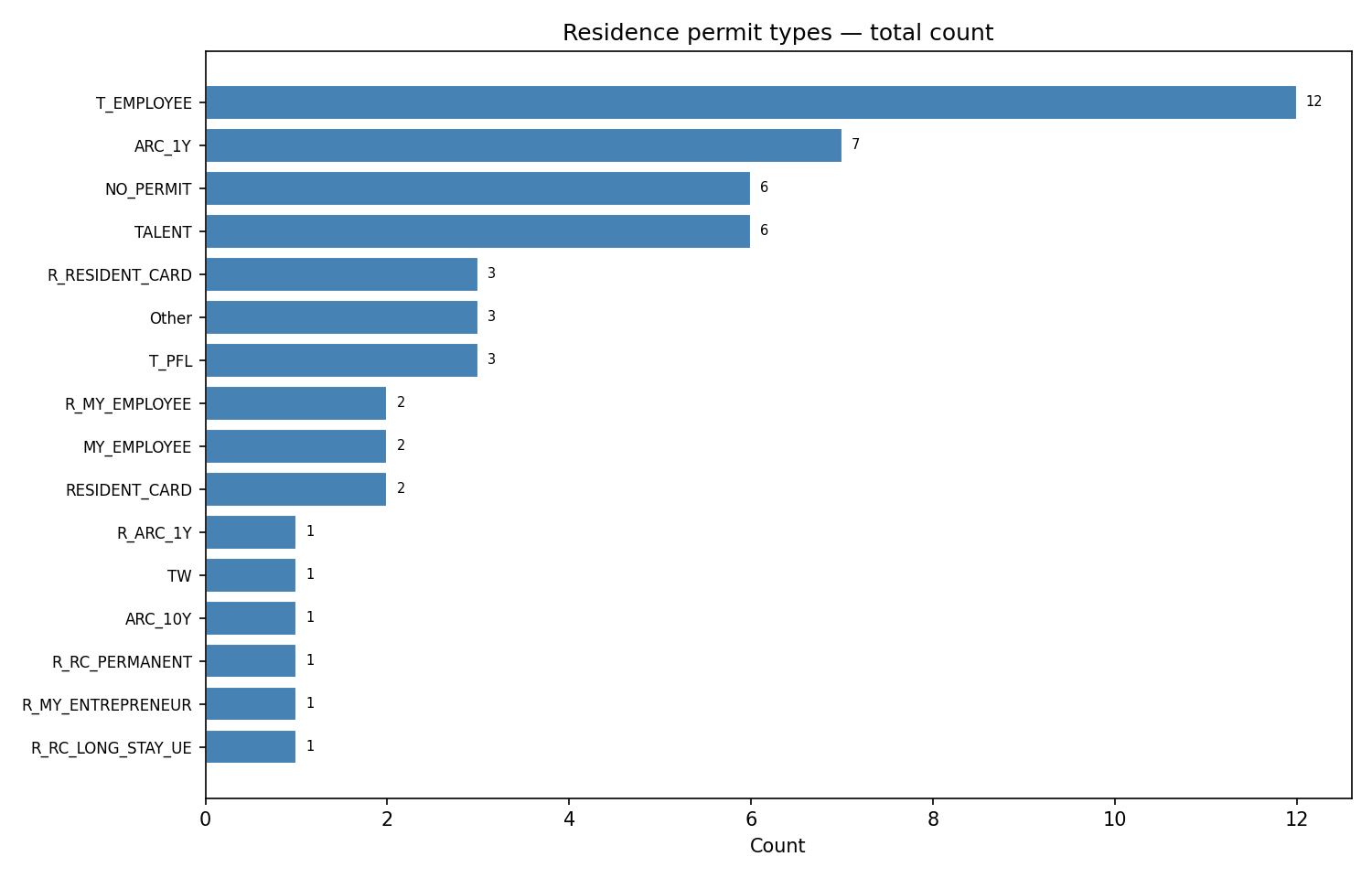}
    \caption{Distribution of recommended permit types (regardless of renewal status)}
    \label{fig:stats_permit_types}
\end{figure}

Figure~\ref{fig:stats_permit_types} reflects the data generation choices: the temporary work permit is the most frequent, consistent with the deliberate oversampling of first-time applications. The one-year Algerian residence certificate ranks second, owing to the strong representation of Algerian nationals and its eligibility for both first-time and renewal applications. Profiles requiring no permit follow, reflecting the prevalence of EU nationals, with the talent card also notably represented.

\begin{figure}[ht]
    \centering
    \includegraphics[width=\linewidth,trim=0 0 0 0.9cm, clip]{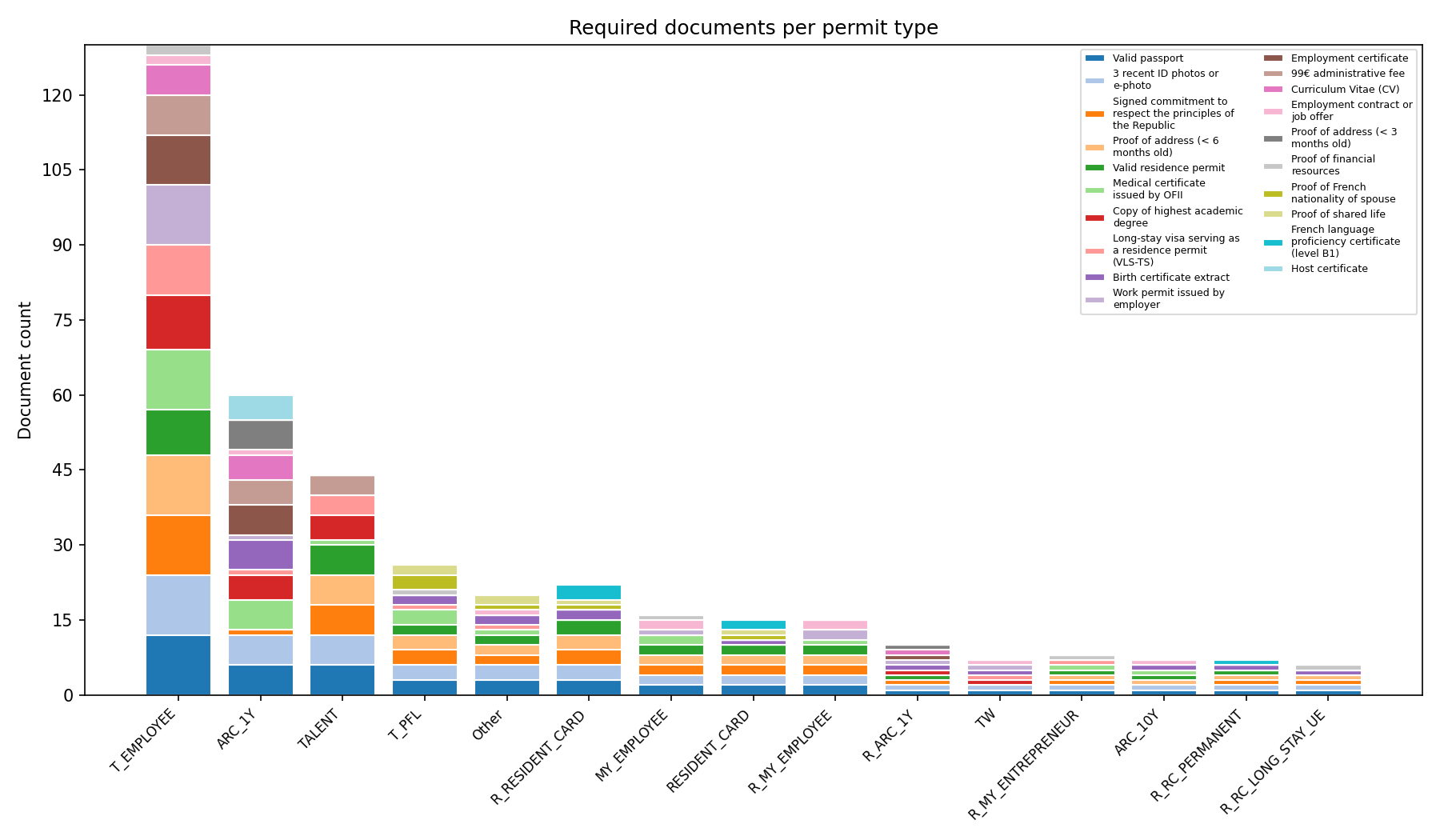}
    \caption{Distribution of recommended documents per permit type (regardless of renewal status)}
    \label{fig:stats_documents}
\end{figure}

\begin{figure}[ht]
    \centering
    \includegraphics[width=\linewidth,trim=0 0 0 0.9cm, clip]{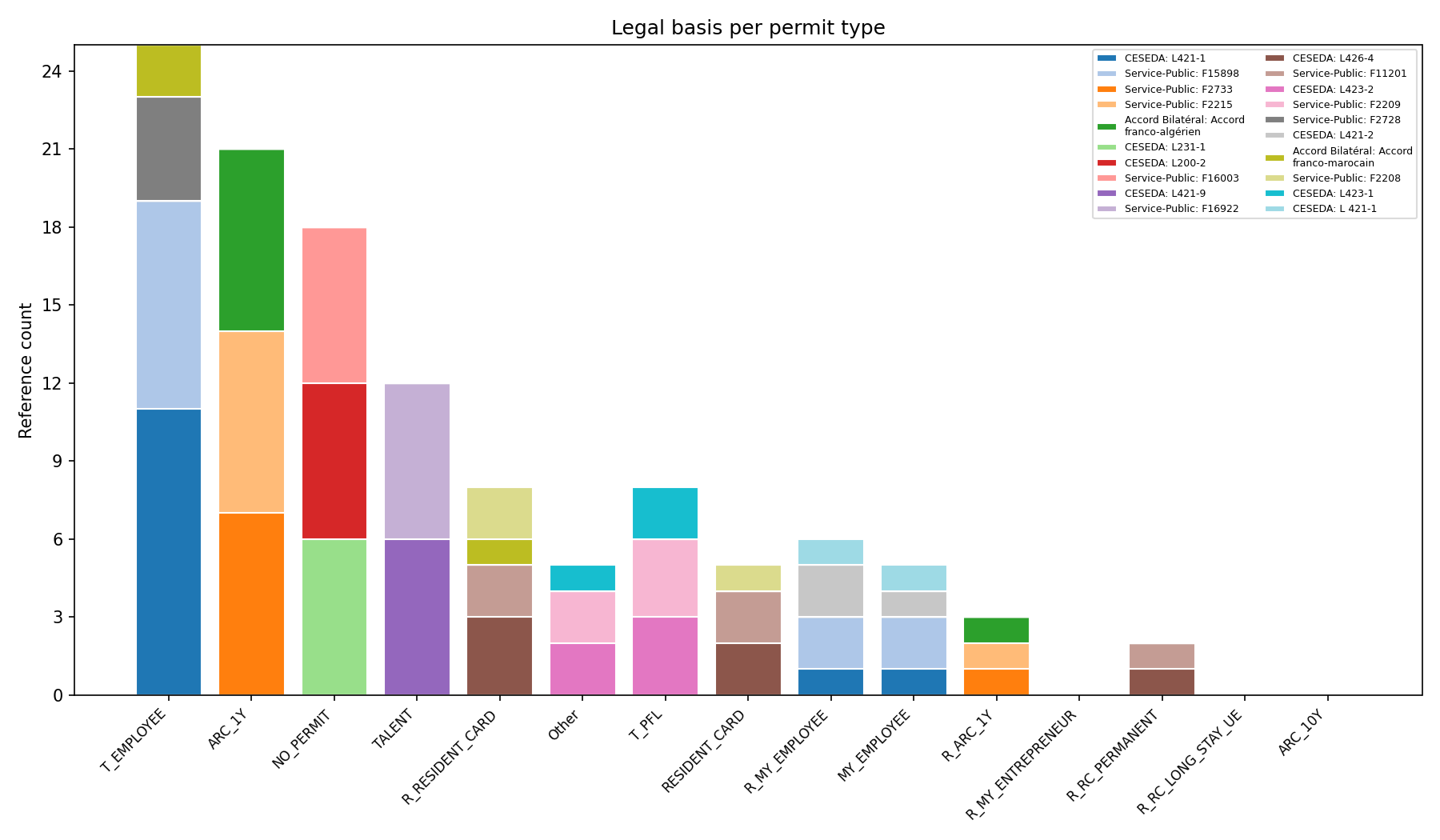}
    \caption{Legal basis distribution per permit type (regardless of renewal status)}
    \label{fig:stats_legal_basis}
\end{figure}

Figure~\ref{fig:stats_documents} reflects permit-type diversity: while most profiles share common documents (passport, identity photo, signed republican commitment, OFII medical certificate), certain permit types impose specific requirements: the Algerian residence certificate, for instance, requires proof of address less than three months old rather than six, along with a host certificate for the visa. Visa requirements further contribute to document variability across profiles. This diversity repercutates on the legal basis, as can be seen in Figure \ref{fig:stats_legal_basis}.

\section{Normalisation and matching hypotheses for the evaluation protocol}
\label{app:normalisation}
\subsection{Permit matching details}
\label{app:permit-matching}

Prior to matching, generated permits are normalised through case folding, accent removal, removal of parenthetical content, and suppression of legal reference strings. When a prediction contains multiple alternatives separated by ``or'', only the first is retained, reflecting the expectation that the system should provide a single primary recommendation.

For EU/EEA/Swiss nationals, predictions mapped to \texttt{UE\_CITIZEN} are treated as equivalent to \texttt{NO\_PERMIT}. Both recommendations correctly reflect the free-movement regime and the absence of a legal obligation to obtain a residence permit. The gold standard nevertheless favours \texttt{NO\_PERMIT}, as this formulation provides the most direct and operational guidance for HR practitioners and applicants.

\subsection{Document matching details}
\label{app:doc-matching}

\paragraph{Canonical layer.}
The rule-based canonical layer resolves frequent surface-form variants prior to embedding-based matching. Covered categories include identity photos (e.g.\ \emph{3 photos d'identité}, \emph{e-photo avec code}), proofs of address, passports and travel documents, OFII medical certificates, work authorisations, and marriage certificates with or without transcription clauses.
\paragraph{Threshold calibration.}
The cosine similarity threshold for BGE-M3 embedding matching was calibrated over ${0.65, 0.70, 0.75, 0.80}$ on the full evaluation subset. Lower thresholds increased recall but frequently matched semantically related yet administratively distinct documents. We therefore retained the most conservative setting ($0.80$), prioritising precision over recall to avoid conflating documents with similar wording but different administrative purposes. This choice is consistent with the evaluation objective, where false matches would artificially inflate document-level performance.

\paragraph{Administrative consistency filters.}
Additional post-matching filters invalidate semantically matched document pairs when critical administrative constraints disagree. Two filters are applied. First, euro-denominated amounts must match exactly, preventing confusions between fee-related documents of different denominations (e.g.\ a €99 visa tax versus a €225 residence permit fee). Second, proofs of address containing explicit validity requirements must agree on the duration constraint (e.g.\ less than 3 months versus less than 6 months). These constraints are frequently overlooked by embedding similarity and are therefore enforced explicitly during matching.

\section{Results}
\label{app:results}

\subsection{Generation prompt} \label{app:prompt}

The prompt comprises four functional blocks: (i) a \emph{system role} and source hierarchy, (ii) \emph{critical rules} encoding known legal special cases, (iii) an ordered \emph{task list} scaffolding the reasoning, and (iv) a strict \emph{output specification}.

Three design choices are central. First, legal sources are ranked by authority (bilateral agreements, then the CESEDA, the Employment Act, and the administrative portal Service-Public), so the model resolves conflicts deterministically. Second, hard-coded rules capture cases where naive retrieval is known to fail: free-movement rights for EU/EEA/Swiss nationals, the precedence of the 1968 Franco-Algerian Agreement over the CESEDA for Algerian nationals, and the effect of shortage-occupation status on eligibility. Third, the model must abstain whenever no authoritative citation supports a recommendation, discouraging unsupported or hallucinated legal claims.

Accompanying quality controls constrain the citation format (one article per line with its collection and canonical reference, e.g.\ \texttt{L.123-45} for the CESEDA and Employment Act, or the \texttt{F}-prefixed identifier for Service-Public) and require the response to specify exact tax amounts, payment procedures, and the recency requirement for proof of address. The full template appears in Figure~\ref{fig:prompt}.

\begin{figure}[ht!]
\centering
\begin{lstlisting}[style=promptstyle]
You are a French immigration law consultant. I am an HR manager at a firm in {cityOfJob}, {postalCodeOfJob}, France.
From the applicant's profile, recommend the appropriate residence permit type to find relevant legal texts and official eligibility guidelines. Cite specific articles. If no authoritative citation exists, state "legal uncertainty" instead of recommending.
 
Legal sources (priority order):
  1. Bilateral agreements
  2. CESEDA
  3. Code du Travail
  4. Service-Public
 
Critical rules:
  - EU/EEA/Swiss citizens: Check free movement rights first
  - Algerians: Apply Franco-Algerian Agreement (1968) for all substantive rules; CESEDA only for procedural gaps. NOT eligible for CESEDA permit types.
  - Metier en tension: Verify shortage occupation status -- significantly affects eligibility
 
Tasks:
  1. Check EU/EEA/Swiss status -> bilateral agreement France-{nationality} -> CESEDA eligibility articles
  2. Verify metier en tension status
  3. Find salary thresholds (Code du Travail / CESEDA carte Talent)
  4. Determine qualifying permit type(s) and required documents
     (Service-Public / CESEDA)
  5. Recommend the most appropriate permit with legal justification and document list
 
Reason internally. Output only the final result as valid JSON in French, matching the provided output format.

{
  "general_recommendation": "<type of permit recommended>",
  "justification": "<reasoning for why this permit type fits the
                     applicant's profile; no legal citations here>",
  "required_documents": ["<doc1>", "<doc2>", "<doc3>"],
  "legal_citations": ["<citation1>", "<citation2>", "<citation3>"]
}
 
Quality controls:
  - Cite the exact article number and collection (CESEDA, Code du Travail, etc.): format FXXXX for Service-Public, and L./R./D.123-45 for Code du Travail and CESEDA. Only one article per line.
  - Never recommend a permit without a supporting legal citation (article and collection).
  - Always give exact tax amounts and payment procedures for the recommended permit, citing the relevant legal articles.
  - For proof of address, always state the recency requirement.
\end{lstlisting}
\caption{Role-conditioned user prompt. Tokens in braces are substituted at runtime from the applicant profile.}
\label{fig:prompt}
\end{figure}

\subsection{Permit matching} \label{app:permit_match}
Table~\ref{tab:error_analysis} summarizes the main permit recommendation errors
identified through manual inspection of model outputs. The most frequent failure
modes involve confusion between legal regimes, permit families, and permit
durations, as well as errors on administrative attributes such as renewal status
and visa prerequisites. We do, however, observe some differences between
Qwen-27B and Qwen-9B. The 9B model produces more vague or generic
recommendations, and tends to fall back on the most common permit categories for
foreign workers in France (\emph{employee} and Talent), even when a more specific
category would be correct or when no permit is required, as for EU nationals.
Overall, though, the same families of errors appear across both models.

\begin{table}[t]
\centering
\caption{Qualitative error categories observed during permit recommendation evaluation.}
\label{tab:error_analysis}
\begin{tabular}{lp{9cm}}
\toprule
\textbf{Error category} & \textbf{Description} \\
\midrule

Legal regime confusion &
Confusion between permits governed by different legal frameworks, most notably CESEDA permits and residence certificates issued under the Franco-Algerian agreement. \\

Permit family confusion &
Confusion between Talent permits and standard salaried permits, or between employment-based and family-based permits. \\

Duration confusion &
Confusion between temporary (\emph{carte de séjour temporaire}) and multi-year (\emph{carte de séjour pluriannuelle}) permits despite correct identification of the underlying permit family. \\

Renewal omission &
Correct permit category predicted but renewal status omitted or incorrectly inferred. \\

Visa prerequisite omission &
Failure to identify required visa prerequisites despite otherwise correct permit recommendations. \\

Mention-level error &
Incorrect statutory mention despite correct permit family. Such errors were comparatively rare once the permit type was correctly identified. \\

\bottomrule
\end{tabular}
\end{table}

\subsection{Document matching results} \label{sec:doc-errors}

Table~\ref{tab:doc_f1} reports document-set matching performance under the strict full permit match filter. As discussed in the main paper, document evaluation is conditioned on correct permit recommendation and should therefore be interpreted as a downstream assessment of document retrieval quality rather than an end-to-end system score.

To better characterise residual failures, Table~\ref{tab:doc_error_typology} reports the document categories most frequently omitted by each system. The results show that retrieval substantially improves recovery of several permit-specific requirements, but some procedural documents remain frequently missing even when the correct permit category has been identified.

\begin{table}[t]
\centering
\caption{Document-set matching performance under the strict full permit match filter at cosine similarity threshold 0.80.
The strict subset contains 9 evaluated profiles for Qwen-27B in the LLM-alone setting, 12 for Qwen-27B with Dense RAG, 5 for Qwen-9B in the LLM-alone setting, and 8 for Qwen-9B with Dense RAG.}
\label{tab:doc_f1}
\begin{tabular}{llrrr}
\toprule
\textbf{Model} & \textbf{Condition} & \textbf{Precision} & \textbf{Recall} & \textbf{F1} \\
\midrule
Qwen 27B \hspace{.5cm} & LLM-alone \hspace{.5cm} & 28.3\% & 21.0\% & 23.9\% \\
 & Dense RAG  & 55.3\% & 39.5\% & 45.3\% \\
 & $\Delta$   & +27.0 pts & +18.5 pts & +21.4 pts \\
\midrule
Qwen 9B  & LLM-alone  & 26.2\% & 21.8\% & 23.5\% \\
  & Dense RAG  & 44.3\% & 31.5\% & 36.4\% \\
  & $\Delta$   & +18.2 pts & +9.7 pts & +12.9 pts \\
\bottomrule
\end{tabular}
\end{table}

\begin{table}[t]
\centering
\caption{Most frequently missing document categories under the strict
\texttt{full\_permit\_match} subset at cosine threshold 0.80.
Each cell reports the number of evaluated profiles for which at least one
gold-standard document in the category was not recovered.}
\label{tab:doc_error_typology}
\begin{tabular}{lcc}
\toprule
\textbf{Document} &
\textbf{LLM-alone} &
\textbf{Dense RAG} \\
\midrule
\multicolumn{3}{l}{\textbf{Qwen-27B} ($n=9$ LLM-alone; $n=12$ Dense RAG)} \\
Signed commitment to respect the principles of the French Republic & 9/9 & 11/12 \\
Administrative fee receipt / tax stamp & 9/9 & 11/12 \\
Proof of address (less than 6 months) & 8/9 & 4/12 \\
Degree copy / diploma evidence & 6/9 & 9/12 \\
\midrule
\multicolumn{3}{l}{\textbf{Qwen-9B} ($n=5$ LLM-alone; $n=8$ Dense RAG)} \\
Signed commitment to respect the principles of the French Republic & 5/5 & 8/8 \\
Administrative fee receipt / tax stamp & 4/5 & 7/8 \\
Degree copy / diploma evidence & 4/5 & 5/8 \\
Birth certificate / full birth record & 3/5 & 3/8 \\
\bottomrule
\end{tabular}
\end{table}


\subsection{Legal basis matching results}
\label{sec:legal-errors}

Table~\ref{tab:citation_coverage} reports citation coverage under the strict \texttt{full\_permit\_match} filter. Results are broken down by legal source type (CESEDA, Service Public, and bilateral agreements). Coverage scores should be interpreted as conservative lower bounds because the gold standard contains only a minimal set of supporting references.

\begin{table}[t]
\centering
\caption{Citation coverage under the strict full permit match filter, broken down by legal source family. Reported scores correspond to gold-coverage recall and should be interpreted as conservative lower bounds because the gold standard contains only a minimal set of supporting references. The strict subset contains 13 evaluated profiles for Qwen-27B in the LLM-alone setting, 18 for Qwen-27B with Dense RAG, 5 for Qwen-9B in the LLM-alone setting, and 13 for Qwen-9B with Dense RAG.}
\label{tab:citation_coverage}
\begin{tabular}{lrr}
\toprule
\textbf{Metric} \hspace{1.5cm} & \textbf{LLM-alone} & \hspace{1.5cm} \textbf{Dense RAG} \\
\midrule
\multicolumn{3}{c}{\textbf{Qwen 27B}} \\
Overall                & 7.7\%           & 46.6\% \\
Service-Public         & 0/14 (0.0\%)    & 11/20 (55.0\%) \\
CESEDA                 & 2/19 (10.5\%)   & 8/28 (28.6\%) \\
Bilateral              & 0/2 (0.0\%)     & 4/4 (100.0\%) \\
\midrule
\multicolumn{3}{c}{\textbf{Qwen 9B}} \\
Overall                & 10.0\%          & 13.1\% \\
Service-Public         & 0/5 (0.0\%)     & 1/14 (7.1\%) \\
CESEDA                 & 1/6 (16.7\%)    & 5/22 (22.7\%) \\
Bilateral              & 0/1 (0.0\%)     & 0/2 (0.0\%) \\
\bottomrule
\end{tabular}
\end{table}


\end{document}